# Performance of prior event rate ratio method in the presence of differential mortality or dropout

Running title: Prior event rate ratio and differential censoring

Yin Bun Cheung [1,2,3], Xiangmei Ma [1]

1. Centre for Biomedical Data Science, Duke-NUS Medical School, National University of Singapore, 8 College Road, Singapore 169857
2. Programme in Health Services Research and Population Health, Duke-NUS Medical School, National University of Singapore, 8 College Road, Singapore 169857
3. Center for Child, Adolescent and Maternal Health Research, Tampere University, Tampere, FIN-33014, Finland.

**ORCID**

Yin Bun Cheung https://orcid.org/0000-0003-0517-7625

Xiangmei Ma https://orcid.org/0000-0001-6526-1226

* Corresponding author:

Professor Yin Bun Cheung, Duke-NUS Medical School, 8 College Road, Singapore 169857

Email: yinbun.cheung@duke-nus.edu.sg

Pre-print: This work is posted at arXiv preprint server (arXiv:2505.20757, version 2.0)

Funding: National Medical Research Council, Singapore (MOH-001487)

**ABSTRACT**

**Purpose:** Prior event rate ratio (PERR) method was proposed to control for measured or unmeasured confounders in real-world evaluation of effectiveness and safety of medical treatments using electronic medical records data. A widely cited simulation study showed that PERR estimate of treatment effect was biased in the presence of differential morality/dropout. However, the study only considered one specific PERR estimator of treatment effect and one specific scenario of differential mortality/dropout. To enhance understanding of the method, we replicated and extended the simulation to consider an alternative PERR estimator and multiple scenarios.

**Methods:** Simulation studies were performed with varying rate of mortality/dropout, including the scenario in the previous study in which mortality/dropout was simultaneously influenced by treatment, confounder and prior event and scenarios that differed in the determinants of mortality/dropout. In addition to the PERR estimator used in the previous study ($\text{PERR}_{\text{Prev}}$) that involved data form both completers and non-completers, we also evaluated an alternative PERR estimator ($\text{PERR}_{\text{Comp}}$) that used data only from completers.

**Results:** The bias of $\text{PERR}_{\text{Prev}}$ in the previously considered mortality/dropout scenario was replicated. Bias of $\text{PERR}_{\text{Comp}}$ was only about one-third in magnitude as compared to that of $\text{PERR}_{\text{Prev}}$ in this scenario. Furthermore, $\text{PERR}_{\text{Prev}}$ did but $\text{PERR}_{\text{Comp}}$ did not give biased estimates of treatment effect in scenarios that mortality/dropout was influenced by treatment or confounder but not prior event.

**Conclusion:** The PERR is better seen as a methodological framework within which there is more than one way to operationalize the estimation. Its performance depends on the specific operationalization. $\text{PERR}_{\text{Comp}}$ provides unbiased estimates unless mortality/dropout is affected by prior event.

## Introduction

Unmeasured confounding is a major challenge in real-world evidence research on the effectiveness and safety of medical treatments. Prior event rate ratio (PERR) was proposed to control the impact of measured or unmeasured confounders.[1,2] It has been applied to the studies of a variety of treatments such as proton pump inhibitors and influenza vaccines.[3,4]

The basic idea of PERR is to partition the person-time of patients who received the treatment concerned during an observation period into two sections: prior to or after the initiation of treatment. Each treated person is matched to at least one control patient who did not receive the treatment during the observation period. The treated person's date of treatment initiation is the index date for partitioning the matched control's person-time into prior and post periods. Then, PERR calculates two estimates of association between outcome and treatment, one prior to and one after the treatment or index date. On the assumption of time-constant confounding, the ratio of the two association estimates cancels out the confounding and provides an estimate of treatment effect. It was originally proposed to study event rate outcomes, where the association estimates were hazard ratios. It has also been applied to binary outcomes, where the association estimate was proportion ratios, or relative risk.[5,6] In any case, the PERR estimate is a ratio of two ratios.

A simulation study demonstrated that PERR estimation of treatment effect on binary outcomes gave biased estimates when mortality/dropout rate was influenced jointly by treatment, confounder and prior event.[6] The study used a PERR estimator that included only completers in the proportion ratio in the numerator but all persons in the proportion ratio in the denominator. This study has been repeatedly cited in both epidemiological and methodological studies.[3,4,7,8]

However, given a dataset, there is more than one way to construct a PERR estimator of treatment effect. There are also multiple pathways that mortality/dropout may be associated with confounder, treatment or prior event. To enhance understanding of the performance of the PERR method, we replicated the previous simulation, evaluated an alternative PERR estimator, and extended the simulation to other scenarios of relationship between mortality/dropout, confounder, treatment or prior event.

## Methods

### PERR estimation of treatment effect

The PERR estimator of treatment effect used in the previous study was:

$$PERR_{Prev} = \frac{E(Y2|X=1 \,\&\, M2=0)/E(Y2|X=0 \,\&\, M2=0)}{E(Y1|X=1)/E(Y1|X=0)} \qquad (1)$$

where X=1 and X=0 denoted groups that ever and never received the treatment of interest during the observation duration, respectively; Y1 and Y2 denoted binary outcomes (0/1) in the prior period and post period, respectively; M2=1 denoted death or dropout in the post period (non-completers) and M2=0 otherwise (completers); E(·) denoted expected value as per standard statistical notation. It was assumed that Y1 was observed for all persons and Y2 was observed only if M2=0. Note that this estimator included only the completers (M2=0) in the numerator but both completers and non-completers in the denominator.

The previous study also evaluated a relative risk estimator:

$$RR = \frac{E(Y2|X=1 \,\&\, M2=0)}{E(Y2|X=0 \,\&\, M2=0)} \qquad (2)$$

In addition to the estimators above, we considered an alternative PERR estimator that included only completers in both the numerator and denominator:

$$PERR_{Comp} = \frac{E(Y2|X=1 \,\&\, M2=0)/E(Y2|X=0 \,\&\, M2=0)}{E(Y1|X=1 \,\&\, M2=0)/E(Y1|X=0 \,\&\, M2=0)} \qquad (3)$$

### Simulation procedures

Uddin et al. considered a simulation scenario of differential mortality/dropout as described in Figure 1, in which all variables are binary.[6] In this framework, mortality/dropout (M2) was directly affected by prior event (Y1), treatment (X) and a confounder (C2). For evaluation of PERR under time-constant confounding, C1 = C2 = C and $\beta_{Y1,C1} = \beta_{Y2,C2}$, where $\beta_{B,A}$ is the regression coefficient representing the effect of A on B.

[Figure 1 about here]

We began with the same simulation scenario – scenario 1 – in the previous study in which M2 was jointly influenced by confounder (C2), treatment (X) and prior event (Y1), as described in Figure 1. The log-binomial regression model parameters for the data generation of scenario 1 are:

$ln(p_{C1}) = ln(0.25)$, and C1 = C2 = C is a time-constant confounder

$ln(p_{Y1}) = -1 + \ln(2)\, C1$

$ln(p_X) = -1.5 + \ln(1.5)\, C1 + \ln(2)\, C2$

$ln(p_{Y2}) = -2 + \ln(2)\, C2 + \ln(2)\, X$

$ln(p_{M2}) = \alpha_{M2} + \ln(4)\, C2 + \ln(1.5)\, X + \ln(1.2)\, Y1$

We extended the simulation to three more scenarios that M2 was influenced by sub-sets of the determinants. In scenario 2, M2 was determined by C2 and Y1. In scenario 3, M2 was determined by C2 and X. In scenario 4, M2 was determined by C2 only. For scenarios 2 to 4, the relevant regression coefficients for X and Y1 in the $ln(p_{M2})$ equation described above are set to zero. In each of the four simulation scenarios, the intercept in the $ln(p_{M2})$ equation, $\alpha_{M2}$, is set such that mortality/dropout rate equals 5%, 10%, 15% or 20%, or M2 is set to zero for all persons for 0% mortality/dropout.

We evaluated the performance of $PERR_{Prev}$, $PERR_{Comp}$ and RR as defined above. As in the previous study, we varied the mortality/dropout rate from 0% to 20% and used 10,000 replicates in each scenario and a sample size of 100,000 persons in each replicate. For each scenario and mortality/dropout level, the mean and 2.5$^{th}$ and 97.5$^{th}$ percentiles of the 10,000 estimates of each estimator were presented. The true effect of X on Y2 is RR=2. That is, other factors being held constant, the probability of Y2=1 was doubled if X=1 as compared to X=0. (Stata codes for the simulation are available at https://github.com/cheungyb/PERR-MD)

## Results

Figure 2 shows the simulation results. In scenario 1 (upper-left panel), we observed the previously reported under-estimation bias of $PERR_{Prev}$. The magnitude increased as the mortality/dropout rate increased. In contrast, $PERR_{Comp}$ was practically unbiased when mortality/dropout rate was below 10%. When it was above 10%, $PERR_{Comp}$ over-estimated the treatment effect. At a mortality/dropout rate of 20%, the average $PERR_{Comp}$ and $PERR_{Prev}$ estimates were 2.05 and 1.83, respectively, as compared to the true RR=2. The (absolute) magnitude of bias in $PERR_{Comp}$ was only about one-third that of $PERR_{Prev}$. The RR estimate was substantially confounded when there was no mortality/dropout. The over-estimation due to confounding was partially cancelled out by the bias arising from differential mortality/dropout as the latter rate increased.

[Figure 2 about here]

In scenario 2 (upper-right panel), where M2 was affected by both confounder and prior event but not treatment, a somewhat similar pattern of bias was seen, but the magnitude was milder. At 20% mortality/dropout, the average $PERR_{Comp}$, $PERR_{Prev}$ and RR estimates were 2.02 and 1.91 and 2.55, respectively.

In scenarios 3 and 4, M2 was not influenced by Y1, even though they were associated due to shared confounding. $PERR_{Prev}$ and RR showed a pattern of bias similar to scenarios 1 and 2. In contrast, $PERR_{Comp}$ provided accurate estimates of the treatment effect in both scenarios regardless of morality/dropout rate.

## Discussion

The control of measured and unmeasured confounding is an important issue in real-world evaluation of medical treatments using observational data such as routinely collected electronic medical records. PERR has been proposed to deal with the challenge and it has been applied to the studies of a variety of treatments. There is a concern about the validity of PERR in the presence of differential mortality or dropout. A simulation study on this issue has been repeatedly cited as the source of this concern.[6]

We propose that PERR method is better seen as a methodological framework within which there is more than one way to operationalize the estimation. Its performance depends on the specific operationalization. Different estimators may be constructed to estimate treatment effect within the PERR framework. Furthermore, mortality/dropout can be associated with outcomes due only to shared confounders. It does not necessarily involve treatment or prior event being the determinants. Defining $PERR_{Comp}$ as the PERR treatment effect estimator using only data from completers, we have shown that PERR based on this estimator provided accurate estimates when mortality/dropout was not influenced by prior event. When mortality/dropout was affected by prior event, its bias was much milder than the previously considered estimator.

## Conclusion

Differential mortality/drop-out does not necessarily induce bias in PERR estimation of treatment effect. It depends on the specific choice of treatment effect estimator. The present findings contribute to better understanding of the PERR method as applied to binary outcomes. The performance of PERR framework and specifications within this framework for event rate outcomes in the presence of differential mortality/dropout requires further investigation.

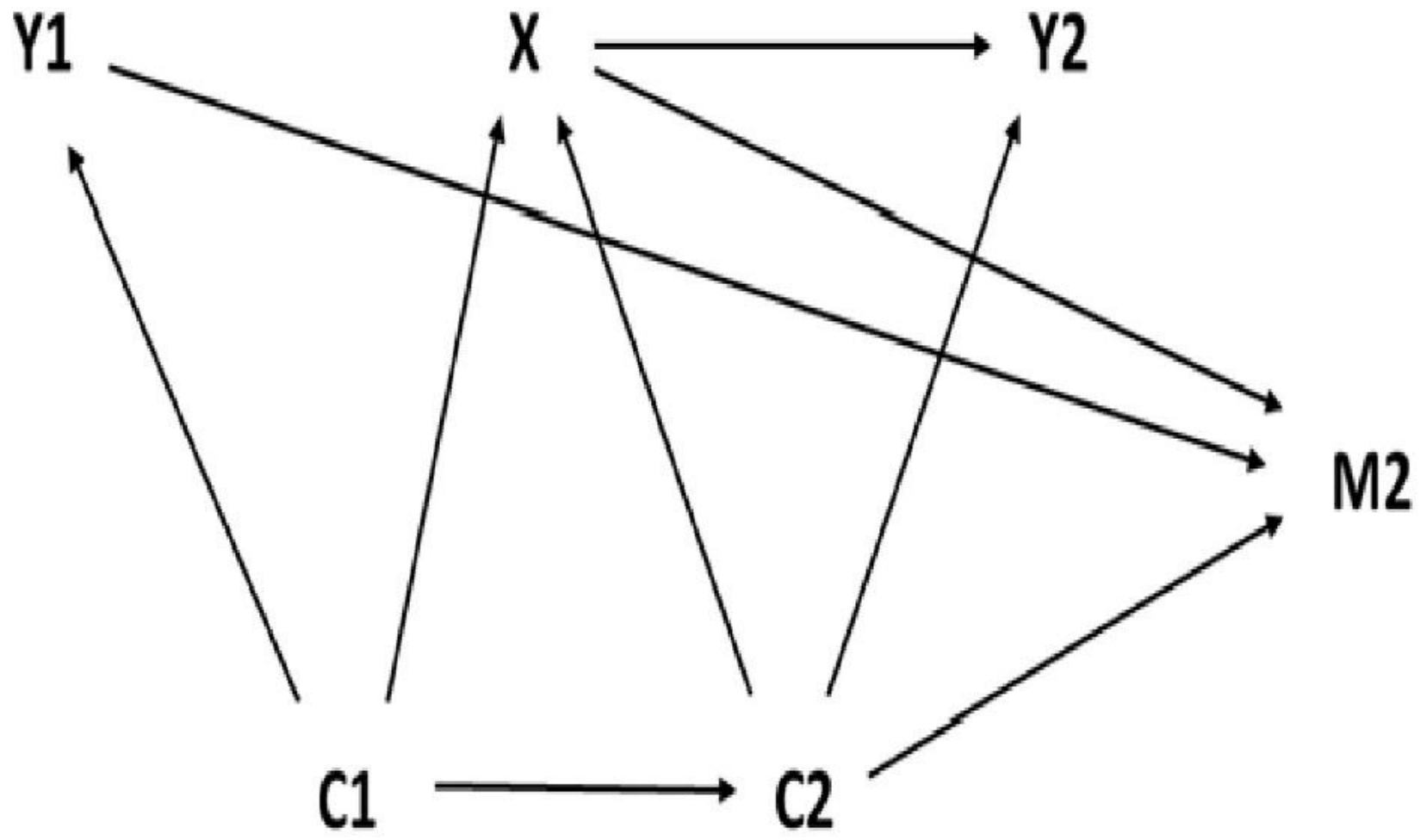

**Figure 1** Simulation framework.

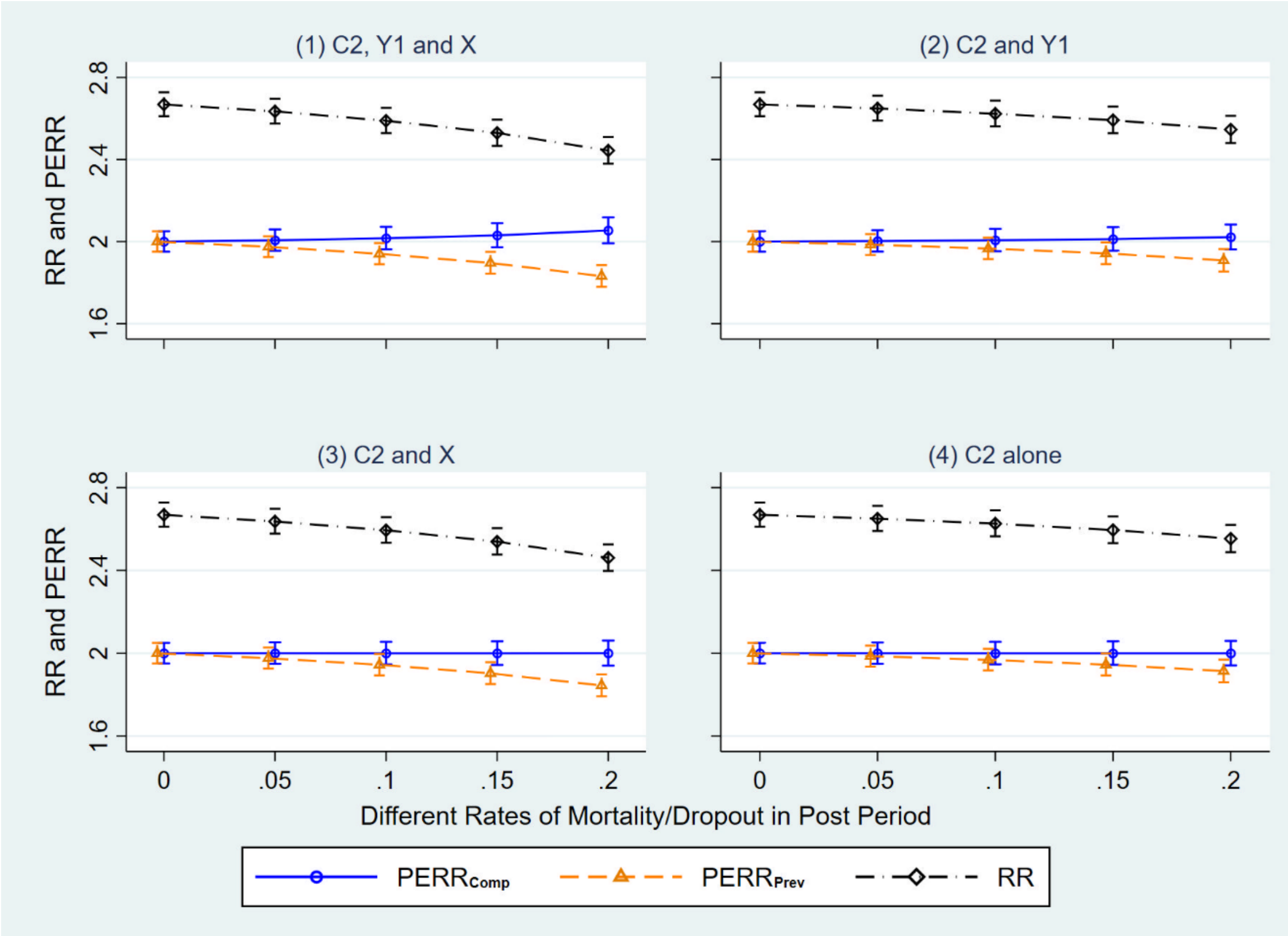

**Figure 2** Simulation results on the performance of treatment effect estimators when M2 depended on: (1) C2, Y1 and X, (2) C2 and Y1, (3) C2 and X, and (4) C2 alone. Markers are mean estimates; lower and upper horizontal bars indicate 2.5th and 97.5th percentiles of estimates, respectively. For visual clarity, the horizontal positions of $PERR_{Comp}$ and $PERR_{Prev}$ are shifted to the right and left by 0.003, respectively.

**Ethics statement**

This is a simulation study and does not require IRB approval.

**Data availability statement**

Stata codes for the simulation are available at https://github.com/cheungyb/PERR-MD.

**Competing interest**

The authors declare no potential conflicts of interest.

**Funding**

This work was supported by the National Medical Research Council, Singapore (MOH-001487).

**Disclaimer**

Any opinions, findings and conclusions or recommendations expressed in this material are those of the authors and do not reflect the views of Ministry of Health / National Medical Research Council, Singapore.